# A BENCHMARK TO SELECT DATA MINING BASED CLASSIFICATION ALGORITHMS FOR BUSINESS INTELLIGENCE AND DECISION SUPPORT SYSTEMS

Pardeep Kumar[1] , Nitin[1] and Vivek Kumar Sehgal[1] and Durg Singh Chauhan[2]

[1]Department of Computer Science & Engineering and Information Technology,
Jaypee University of Information Technology, Waknaghat, Solan, Himachal Pradesh, India
`pardeepkumarkhokhar@gmail.com`, {`delnitin,vivekseh`}`@ieee.org`
[2]Institute of Technology , Banaras Hindu University, Banaras, U.P., India
Currently with Uttarakhand Technical University, Deharadun, Uttarakhand, India
`pdschauhan@acm.org`

## ABSTRACT

*In today's business scenario, we percept major changes in how managers use computerized support in making decisions. As more number of decision-makers use computerized support in decision making, decision support systems (DSS) is developing from its starting as a personal support tool and is becoming the common resource in an organization. DSS serve the management, operations, and planning levels of an organization and help to make decisions, which may be rapidly changing and not easily specified in advance. Data mining has a vital role to extract important information to help in decision making of a decision support system. It has been the active field of research in the last two-three decades. Integration of data mining and decision support systems (DSS) can lead to the improved performance and can enable the tackling of new types of problems. Artificial Intelligence methods are improving the quality of decision support, and have become embedded in many applications ranges from ant locking automobile brakes to these days interactive search engines. It provides various machine learning techniques to support data mining. The classification is one of the main and valuable tasks of data mining. Several types of classification algorithms have been suggested, tested and compared to determine the future trends based on unseen data. There has been no single algorithm found to be superior over all others for all data sets. Various issues such as predictive accuracy, training time to build the model, robustness and scalability must be considered and can have tradeoffs, further complex the quest for an overall superior method. The objective of this paper is to compare various classification algorithms that have been frequently used in data mining for decision support systems. Three decision trees based algorithms, one artificial neural network, one statistical, one support vector machines with and without adaboost and one clustering algorithm are tested and compared on four datasets from different domains in terms of predictive accuracy, error rate, classification index, comprehensibility and training time. Experimental results demonstrate that Genetic Algorithm (GA) and support vector machines based algorithms are better in terms of predictive accuracy. Former shows highest comprehensibility but is slower than later. From the decision tree based algorithms, QUEST produces trees with lesser breadth and depth showing more comprehensibility. This research work shows that GA based algorithm is more powerful algorithm and shall be the first choice of organizations for their decision support systems. SVM without adaboost shall be the first choice in context of speed and predictive accuracy. Adaboost improves the accuracy of SVM but on the cost of large training time.*

# 1. INTRODUCTION

Despite successes in recent years in the area of large scale database design, we are still challenged by the difficulties associated with unlocking the data we need and removing it from the cavernous databases in which it resides. In addition, we are becoming increasingly aware of the hidden treasure trove of new knowledge quietly residing in our data and face considerable frustrations when we attempt to get it. Such a never ending cycle of data creation, storage and problem in its access and analysis has resulted in the implementation of new and efficient tools to help us in handling such challenge. There is adequate amount of observed evidence that human judgment and decision making can be too far from the best, and it becomes worst even further with complexity and stress. Various areas like statistics theory, economics and operations research have different methods for selecting choices. In the last decades, such methods, with a variety of techniques coming from information theory, psychology, and artificial intelligence, have been developed in the form of computer software or programs, either as independent tools or as combined computing environments for complex decision making. Such programs are named as decision support systems (DSSs). According to Gorry and Morton , "A DSS is an interactive computer based system that helps decision makers utilizes data and models to solve unstructured problems"[1]. It is a computer-based support system for management decision makers who deal with semi structured problems. The idea of DSS is wide, and its definitions may vary from author's perception in literature.

Data mining is defined as the extraction of hidden knowledge, exceptional patterns and new findings from huge databases. Data mining is considered as the key step of a detailed process called *Knowledge Discovery in Databases (KDD)* which is defined as the non – trivial process of identifying valid, novel, and ultimately understandable patterns in large databases [2].

The bridging of data mining and decision support has a significant impact on the developments of both fields, largely by improving approaches for problem solving in real settings, enabling the fusion of knowledge from experts and knowledge extracted from data, and consequently enabling the successful solution of new types of problems[3]. Mladenic, D. and Lavrac,N have done excellent work for the integration of these two research areas in *SolEuNet* project[4-5].

Classification is one of the important tasks of data mining. There is lot of research going in the machine learning and statistics communities on algorithms for classification algorithms. The conventional models used for classification are decision trees, neural network, statistical and clustering techniques. SVM is the recent development and widely used these days for classification.

There is a project, called the STATLOG project, tests and compares the predictive accuracy of various decision tree classification algorithms against some non-decision tree classification algorithms on a huge number of datasets [6]. This project discovers that no classification algorithm is exactly most accurate over the datasets tested and various algorithms possess sufficient accuracy. Earlier comparative studies put emphasis on the predictive accuracy of classification algorithms; other factors like comprehensibility are also becoming important. Breslow and Aha have surveyed methods of decision tree simplification to improve their comprehensibility [7]. Brodley and Utgoff , Brown, Corruble, and Pittard, Curram and Mingers, and Shavlik, Mooney and Towell have also tested and compared various classification algorithms on the datasets from varying domains[8-11]. Roger J Marshall, P Baladi, S Brunak, Y Chauvin , C.A Anderson have done latest work in the selection of classification algorithms based on various important parameters like misclassification rates and types of attributes at hand[27-28, 30]. Saroj and K.K Bhardwaj have done excellent work to exploit the capability of genetic algorithms to discover information from huge data repositories [31]. Bikash Kanti Sarkar, Shib Sankar

Sana, and Kripasindhu Chaudhuri have made accuracy based learning classification system where C4.5 and GA capabilities has been expored in classification domain [32].

This paper compares three decision trees (CHAID, QUEST and C4.5), one neural network (Back Propagation), one statistical (Logistic regression), one support vector machine(LibSVM and AdaboostM1-SVM) with and without boosting and one clustering algorithm (k-means). These algorithms are tested on four datasets (Mushroom,Vote,Nursery and Credit) that are taken from the University of California, Irvine, Repository of Machine Learning Databases (UCI) [14]. Further, genetic algorithm is tested on all the data sets. Here, section 2 briefly describes the algorithms and section 3 describes some background to the datasets and experimental setup, and Section 4 shows the result. Conclusion is given in section 5.

## 2. THE ALGORITHMS

### 2.1. Decision Trees

#### 2.1.1. CHAID

Such a decision tree based algorithm is based on a statistical approach called chi square test. CHAID acronym expansion is chi square automatic interaction detection. It is different from other decision tree algorithms in the sense of attribute selection measure for tree formation. It uses chi square test to choose best split instead of information gain (reduction in entropy) as in C 4.5 during tree generation. It is having automatically constructing a contingency table, and checking out statistical importance of the proportions. The most important correlations among attributes are used to form the shape of a tree. It includes (i) computing the expected frequencies (ii) application of the chi square formula (iii) calculate the degree of freedom and (iv) using the chi square table. Attributes that are not statistically significant are merged during tree formation and those attributes that are statistical significant become nodes in the tree. More detail can be found at [7, 16].

#### 2.1.2. QUEST

The acronym QUEST stands for Quick, Unbiased, Efficient Statistical Tree. It is a binary classification algorithm for constructing decision trees. A major motivation in its development was to reduce the processing time required for large C&RT (classification & regression tree) analyses with either many variables or many cases. Secondly QUEST was to decrease the trend found in classification tree algorithms to give priority to attributes that permit more splits; i.e. continuous attributes or those with multiple categories. QUEST uses a sequence of rules, based on significance tests, to evaluate the predictor variables at a node. For selection purposes, as little as a single test may need to be performed on each predictor at a node. Unlike C&RT, all splits are not examined, and unlike C&RT and CHAID, category combinations are not tested when evaluating a predictor for selection. This speeds the analysis. Splits are determined by running quadratic discriminate analysis using the selected predictor on groups formed by the target categories. This method again results in a speed improvement over exhaustive search (C&RT) to determine the optimal split [6, 17].

#### 2.1.3. C 4.5

C 4.5 algorithm uses information gain measure to select the test attribute at each node in the tree. Such a parameter is termed as an attribute selection measure or a measure of goodness of split. The attribute having largest information gain (or maximum entropy reduction) is selected as the test attribute for the current node. Such attribute reduces the randomness or information required to classify the tuples in the generated partitions and represents the least randomness or impurity in terms of classification in such partitions. Such an approach based on information theory reduces the expected number of comparisons or tests required to classify an object and assure that a tuple (but not necessarily the simplest) tree is found. More detail of the algorithm can be found at [7, 16].

## 2.2. Neural Network

**Back Propagation:** It is an example of artificial neural network. It learns by recursively processing a set of training tuples, comparing the network's observed output for each tuple with the actual known class attribute value. For each training tuple, the weights are edited so as to reduce the mean squared error between the network's output and the actual class label or value. These changes are propagated in backward direction, i.e. from the output layer, through each hidden layer down to the first hidden layer. After running the process repetitively, the weights will finally converge, and the training process stops. More detail of this algorithm can be found at [9-13, 15].

## 2.3. Statistical Techniques

**Logistic regression:** It is a technique from statistics for classifying tuples by considering values of input fields. It is applicable for categorical as well as numerical class attributes whereas linear regression requires only numerical class attributes. In this technique, set of equations are generated that link the input field values to the probabilities related with each of the output field categories or classes. Once the model is built, it can be used to calculate probabilities for new tuples. For each tuple, a probability of membership is calculated for each possible output category or class label. The class category or value with the maximum probability is assigned as the predicted output value for that tuple. Probabilities calculation is carried out in this technique by a logistic model equation. More detail of the algorithms can be found at [15,18-21].

## 2.4. Clustering Techniques

**k-means:** The k-means provides a method of cluster analysis. It is used to categorize the tuples of a dataset in different groups based on the similarities. It is an example of supervised machine learning. The class label of the tuples is not known while training. Instead of trying to predict an outcome, k-means tries to uncover patterns in the set of input fields. Tuples are arranged in groups so that tuples of a group or cluster seem to be similar to each other, but tuples in different groups are dissimilar. k-means works by initializing a set of initial cluster centers generated from dataset. The tuples are put in a cluster to which they are most similar. Similarity is calculated by Euclidean formula. Such formula considers the tuple's input field values. After all tuples have been put in clusters, the cluster centers are modified to show the new set of tuples assigned to each cluster. The tuples are then tested again to check whether they should be re located to a different cluster, and the tuple assignment/group repetition process goes until either the highest number of iterations is achieved, or the change between one repetition and the next fails to exceed a predefined threshold. More details of the algorithm can be found at [6,15].

## 2.5. Evolutionary Techniques

**Genetic Algorithm:** Genetic Algorithm (GA) is based on Darwinian natural selection and Mendelian genetics, in which each point in the search space is a string called a chromosome that represents a possible solution. In this approach there is a requirement of a population of chromosomes used to represent a combination of features from the set of features and a function that computes each chromosome's fitness (such a function is called evaluation function or fitness function in literature). The algorithm does an optimization by editing a finite population of chromosomes. In each generation, the GA generates a set of new chromosomes by three core operation known as crossover, inversion and mutation [22-25-26]. The pseudo code for genetic algorithm used in data mining is given below. In this pseudo code, initial population represents encoded production rules. Fitness function is defined in terms of predictive accuracy and comprehensibility given by formula as Fitness function is given by the formula as

*Fitness Function ∝ Prdeictic Accuracy × Comprehensibility*

Pseudo Code-Genetic Algorithm in Data Mining

   *{*

       1.  *Create initial population;*
       2.  *Compute fitness of individuals(binary encoded rules);*
       3.  *REPEAT*
       4.  *Sort rules in decreasing order of Fitness;*
       5.  *Select individuals based on fitness;*
       6.  *Store the sorted rules into CandidateRuleList;*
       7.  *WHILE (CandidateRuleList is not empty) AND (TrainingSet is not empty)*
       8.  *Remove from the TrainingSet the data instances correctly covered by the first rule in CandidateRuleList ;*
       9.  *Remove the first rule from CandidateRuleList and insert it into SelectedRuleList;*
      10. *ENDWHILE*
      11. *Apply genetic operators to selected individuals, creating offspring;*
      12. *Compute fitness of offspring;*
      13. *Update the current population;*
      14. *UNTILL (stopping criteria)*

   *}*

## 2.6. Support Vector Machines

**LibSVM:** A Support Vector Machines (SVM) is an algorithm for the classification of both linear and non-linear data. It maps the original data in large dimensions, from where it can find a hyper plane for division of the data using important training samples referred as support vectors. Support vector machines are based on a principle from computational learning theory which is called as structural risk minimization principle. The concept of this principle is to search a hypothesis h for which we can assure the lowest true error. Such error of h is the probability that h will make an error on an unseen and arbitrarily selected test example. A maximum limit can be used to link the true error of a hypothesis h with the error of h on the training set and the difficulty of H (measured by VC-Dimension), the hypothesis space having h. Support vector machines find the hypothesis h which (nearly) reduces this limit on the true error by properly handling the VC-Dimension of H. SVM are global learners. Basically SVM learn linear threshold function. With selection of a proper kernel function, they can learn polynomial classifiers, radial basic function (RBF) networks, and 3 layer sigmoid neural networks. One vital characteristic of SVM is the capacity to learn can be independent of the dimensions of the feature space. SVM calculates the difficulty of hypotheses based on the margin with which they separate the data, not the number of features. So we can say that even in the presence of different features, if data is separable with a large margin using functions from the hypothesis space. The same margin also suggests a heuristic for selecting good parameter settings for the learner. The same margin argument also provides a heuristic for selecting better parameter settings for the learner as the kernel width in an RBF network [15, 33, 36-38].

LibSVM in WEKA tool[34] simulation environment to illustrate Support Vector Machines [35].

## 2.7. Boosting

**Adaboost:** Adaboost is a popular ensemble boosting algorithm. Let us assume that we would like to increase the accuracy of classification algorithm.  From a dataset D, a data set of d class-labeled samples, $(X_1, y_1), (X_2, y_2),\ldots, (X_d, y_d)$, where $y_i$ is the class label of sample $X_i$. In its start Adaboost put each training samples an equal weight of $\frac{1}{d}$.  Creating k classifiers for the ensemble needs k cycles in the rest of the algorithm. In cycle i, the samples from D are sampled to form a training set, $D_i$, of size d. Sampling with replacement is used i.e. the same sample may be considered more than once. Each sample's of being selected is dependent on its weight. A classifier model, $M_i$, is generated from the training samples of $D_i$. Its error is computed using $D_i$ as a test set. The weights of the training samples are modified as how they were identified with class values. If a sample was correctly classified, its weight is adjusted to a large value. If a sample was incorrectly classified, its weight is adjusted to a small value. A sample's weight shows how difficult it is to classify. We can say that higher the weight, the more often it has been incorrectly classified. Such weights will be used to create the training samples for the classifier in the next round. The basic concept is that when we make a classifier, focus should be more on the misclassified tuples of the previous round. Finally we create a sequence of classifiers complementing each other [39].

AdaboostM1 with LibSVM is used in WEKA [34] simulation environment to illustrate the performance enhancement using boosting.

# 3. EXPERIMENTAL SETUP

## 3.1. Data Sets

There are four datasets used in this research work from real domain. All the datasets are available in UCI machine learning repository [14].

Mushroom: It includes details of hypothetical samples related to 23 species of gilled mushrooms in the Agaricus and Lepiota family. Every species is categories as edible or poisonous.

Vote: This data set includes votes for each of the U.S. House of Representatives Congressmen on the 16 key votes identified by the CQA. The CQA includes nine different variety of votes: voted for, paired for, and announced for , voted against, paired against, and announced against (these three simplified to nay), voted present, voted present to avoid conflict of interest, and did not vote or otherwise make a position known. Democrat and Republic are the two distinct class attribute values.

Nursery: This data set was originally developed to rank applications for nursery schools. It was exercised in the 1980's when there was huge enrollment to the schools in Ljubljana,Slovenia, and the rejected application frequently needed an detailed explanation. This data set is used to predict whether application is rejected or accepted. The final decision depends on occupation of parents and child's nursery,family structure and financial standing, and social and health picture of the family. The class attribute contains five values: not_recom, recommend, very_recom, priority and spec_prior.

Credit: This data set concerns credit card application. Based on the survey on individuals at a Japanese company that grants credit created the dataset. The class field represents positive and negative instances of people who were and were not granted credit. The class attribute is represented by +(Credit granted) and –(credit not granted). All field names and their values have been modified to meaningless representations to maintain privacy of the data. The table for above four data sets is shown below:

Table1. Composition of Data Sets

| S. No | Set | Total Size | Missing Value | Effective Size | Class | Total Attributes |
|---|---|---|---|---|---|---|
| 1. | Mushroom | 8124 | 2115 | 5609 | 2 | 23(Nominal valued) |
| 2. | Vote | 435 | 204 | 231 | 2 | 1(Nominal)+16(Boolean valued) |
| 3. | Nursery | 12960 | 0 | 12960 | 5 | 9(Nominal valued) |
| 4. | Credit | 690 | 39 | 651 | 2 | 1(Boolean)+ 6(Continuous)+ 9(Nominal)Valued |

### 3.2. Parameters for Comparison

1. Predictive accuracy: It is defined as the percentage of correct prediction made by a classification algorithm [2, 15, 25].

2. Error rate: It is defined as the percentage of wrong prediction made by a classification algorithm [6, 17].

3. Training time: It is defined as the time that an algorithm takes to build a model on datasets. Minimum training time is desirable [15, 17].

4. Classification index: It is a term that describes the degree of amount of information (in bits) required to classify class attribute on datasets. Minimum classification is desirable [15].

5. Comprehensibility: It shows degree of simplicity in rule sets obtained after classification. Higher degree of comprehensibility is required. Greater the number of leaf nodes and depth of tree, lesser will be the comprehensibility [11, 15-16].

### 3.3. Implementation

All the non evolutionary approach based algorithms excluding support vector machines and adaboost have been applied to test data sets using Clementine 10.1[40] on a Pentium IV machine with Window XP platform. SVM with and without boosting have been applied to test data sets using WEKA [34] on the same machine. LibSVM and AdaboostM1 are used as SVM and boosting. In case of LibSVM, kernel function is radial basis function and SVM type is C-SVM while rest of the parameters remains default. AdaboostM1 is used as boosting method on LibSVM classifier. Entropy evaluation measure based K & B information statistics measure is used as classification index for LibSVM and AdaboostM1.Evolutionary approach based algorithm have been applied to test data set using GALIB 245 simulator [41] on same machine with Linux platform. All the data mining algorithms have been run 10 times with 10 fold cross validation and average outcome is recorded.

## 4. RESULT

### 4.1. Predictive Accuracy

Table 2 and Figures (1-2) show the predictive accuracy on Mushroom, Vote, Nursery and Credit data sets for different classification algorithms.

Table 2. Predictive accuracy on Mushroom, Vote, Nursery and Credit data sets.

| Data Set<br>Classifier | Mushroom | Vote | Nursery | Credit |
|---|---|---|---|---|
| CHAID | 98.36 % | 99.69% | 91.02 % | 86.78% |
| QUEST | 86.57% | 99.39% | 86.88% | 86.57% |
| C4.5 | 96.0% | 100% | 94.3% | 88.55% |
| Neural N/W | 100% | 92.21% | 92.34% | 86.68% |
| Logistic Regression | 99.9% | 90.93% | 69.0% | 80.44% |
| k-means | 85.29% | 80% | 100% | 52% |
| Genetic Algorithm | 98% | 94% | 97.3% | 96.2% |
| SVM | 100% | 96% | 55% | 86% |
| SVM-ABoostM1 | 100% | 97% | 60% | 95% |

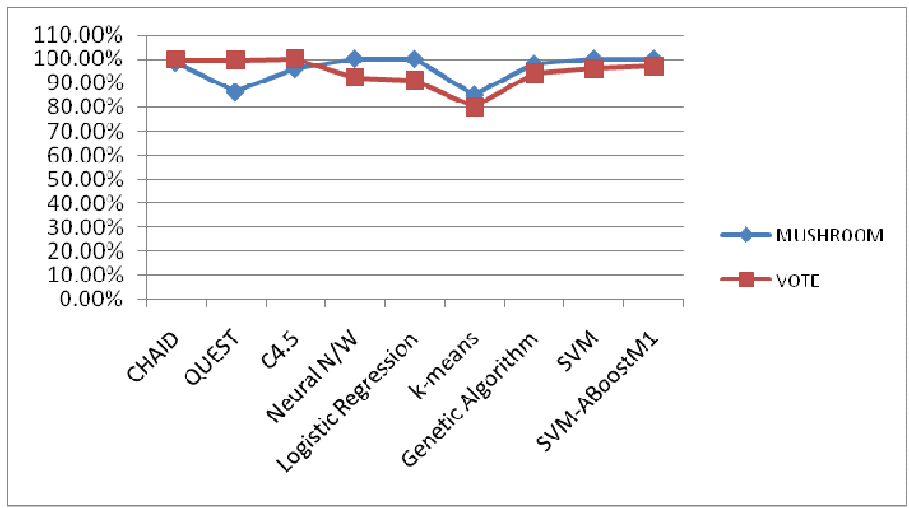

Figure1. Predictive Accuracy on Mushroom and Vote data sets.

Figure1 describes the performance of classification algorithms on Mushroom and Vote data sets in terms of predictive accuracy. SVM, Genetic algorithm, neural network and logistic regression show good predictive accuracy. CHAID and C4.5 occupy the second position but the later shows excellent performance on Vote data set. QUEST and k-means occupy the last position. The former is on the better position in case of Vote data set. Predictive accuracy of SVM increases by applying ABoostM1.

Figure 2 describes the performance of classification algorithms on Nursery and Credit data sets in terms of predictive accuracy. k-means algorithm shows excellent performance on Nursery data set. This might be due to the larger number of distinct values of class attribute. Genetic algorithm shows the excellent performance on both of the data sets. Neural network and CHAID occupy the third position.

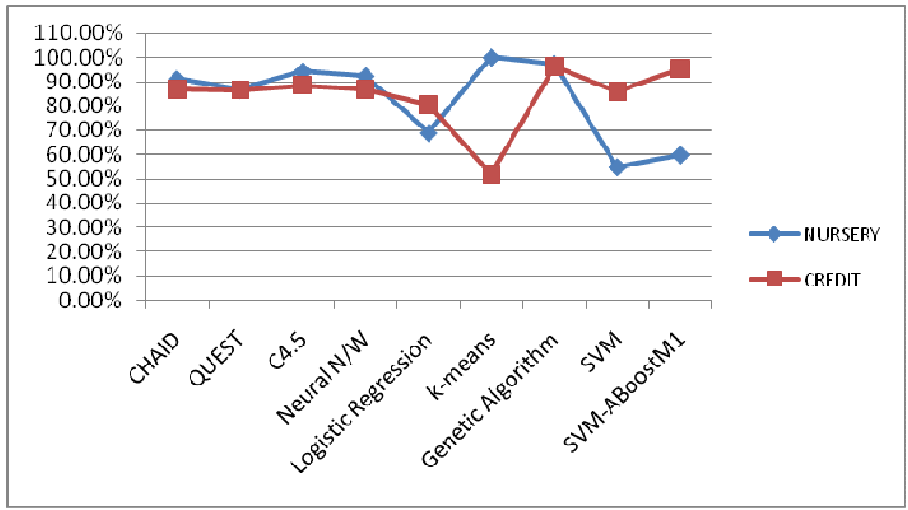

Figure2. Predictive Accuracy on Nursery and Credit data sets.

QUEST occupies the position ahead of k-means in case of Credit data set. The later occupies the last position on Credit data set. SVM performs worst on Nursery data set. This may be due to large number of plane formations required for classification as the large number of class attributes distinct values. Again its performance is enhanced with ABoostM1. It shows accuracy of the order of decision trees on Credit data set. Its performance increases with ABoostM1 and approximately becomes equal to GA on the same data set.

## 4.2. Error Rate

Table 3 shows the error rate on Mushroom, Vote, Nursery and Credit data sets for all classifiers.

Table3. Error Rate on Mushroom, Vote, Nursery and Credit data sets.

| Data Set Classifier | Mushroom | Vote | Nursery | Credit |
|---|---|---|---|---|
| CHAID | 1.6 % | 0.31% | 8.98 % | 13.22% |
| QUEST | 13.13% | 0.61% | 13.12% | 13.43% |
| C4.5 | 4.0% | 0% | 5.7% | 11.45% |
| Neural N/W | 0% | 7.79% | 7.66% | 13.12% |
| Logistic Regression | 0.1% | 9.07% | 31% | 19.56% |
| k-means | 14.71% | 20% | 0% | 48% |
| Genetic Algorithm | 2% | 6% | 2.7% | 3.8% |
| SVM | 0% | 4% | 45% | 14% |
| SVM-ABoostM1 | 0% | 3% | 40% | 5% |

Error rate table is self explanatory as it is equal to 100-predictive accuracy.

### 4.3. Training Time

Table 4 and Figure 3 show the training time (in seconds) on Mushroom, Vote, Nursery and Credit data sets.

Table 4. Training Time on Mushroom, Vote, Nursery and Credit datasets.

| Data Set Classifier | Mushroom | Vote | Nursery | Credit |
|---|---|---|---|---|
| CHAID | 60 | 60 | 60 | 60 |
| QUEST | 60 | 60 | 60 | 60 |
| C4.5 | 60 | 60 | 60 | 60 |
| Neural N/W | 240 | 60 | 60 | 120 |
| Logistic Regression | 120 | 60 | 60 | 60 |
| k-means | 60 | 60 | 60 | 60 |
| Genetic Algorithm | 180 | 60 | 120 | 180 |
| SVM | 32 | 2 | 300 | 2 |
| SVM-ABoostM1 | 300 | 20 | 350 | 42 |

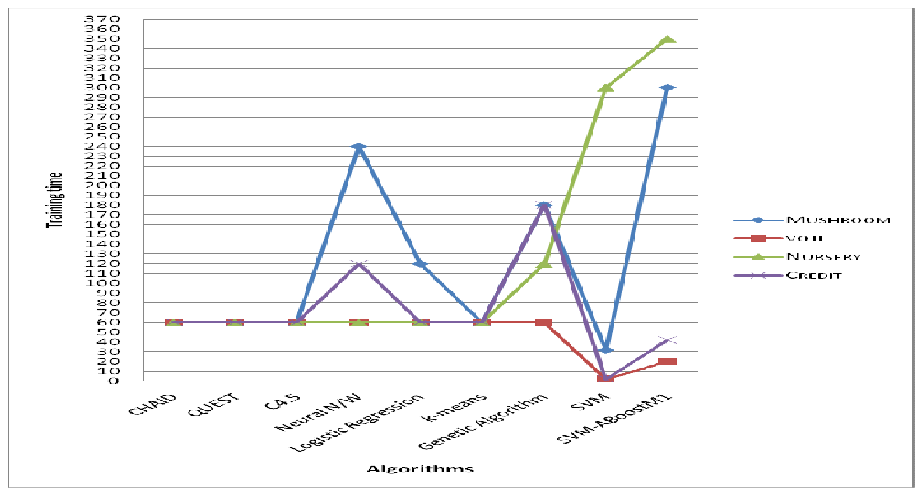

Figure3. Training time on Mushroom, Vote, Nursery and Credit data set.

Figure 3 describes the training time on Mushroom, Vote, Nursery and Credit data sets. Genetic algorithm and neural network are the slowest algorithms on Mushroom and Credit data sets. The former continues the same performance on other data set also. Logistic regression shows the second worst performance.  SVM is fastest one among all the algorithms on all the data sets except nursery. On this data set it is slowest among all algorithms due to large number of distinct values of class attribute present. SVM-ABoostM1 always takes large time due to boosting process behavior on all the data sets.  Rests of the algorithms are fast on both the data sets.

## 4.4. Classification Index

Table 5 and figure (4-7) show the classification index on Mushroom, Vote, Nursery and Credit data sets for all classification algorithms except genetic algorithm.

Table5. Classification Index on Mushroom, Vote, Nursery and Credit datasets.

| Data Set Classifier | Mushroom | Vote | Nursery | Credit |
|---|---|---|---|---|
| CHAID | e=0.481 p=0.955 | d=0.635 r=0.681 | N_R=1.095 P=0.964 S_P=1.046 V_R=3.148 | +=0.563 -=0.541 |
| QUEST | e=0.476 p=0.954 | d=0.502 r=0.752 | N_R=1.094 P=0.862 S_P=0.985 V_R= | +=0.537 -=0.574 |
| C4.5 | e=0.474 p=0.974 | d=0.631 r=0.706 | N_R=1.098 P=1.032 S_P=1.067 V_R=3.407 | +=0.647 -=0.503 |
| Neural N/W | e=0.481 p=0.962 | d=0.494 r=0.733 | N_R=1.089 P=1.032 S_P=1.04 V_R=2.997 | +=0.594 -=0.515 |
| Logistic Regression | e=0.475 p=0.972 | d=0.597 r=0.717 | N_R=1.102 P=0.986 S_P=1.041 V_R=3.379 | +=0.593 -=0.41 |
| K-Mean | Nil | Nil | NIL | Nil |
| SVM | 0.9 | 0.8 | 1.5 | 0.8 |
| SVMAdaboost M1 | 0.9 | 0.9 | 2 | 0.9 |

classification algorithms except genetic algorithm. K & B information statistics measure is used as an indicator of classification index for SVM and SVM-ABoostM1. This parameter is not included in the graph but its value is used for comparison from the table directly.

Note: From figure (4-7), all symbols represent class attribute values of all the datasets.

e=edible, p=poisonous, d=democratic, r=republic, N_R=not recommended, P=priority, S_P=special_priority,S_P=special priority, V_R=very recommended,+=credit granted, -=credit not granted.

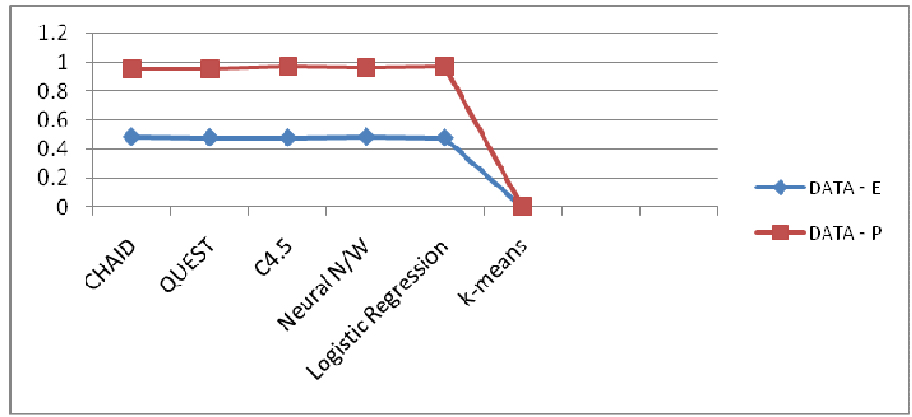

Figure4. Classification Index on Mushroom data set.

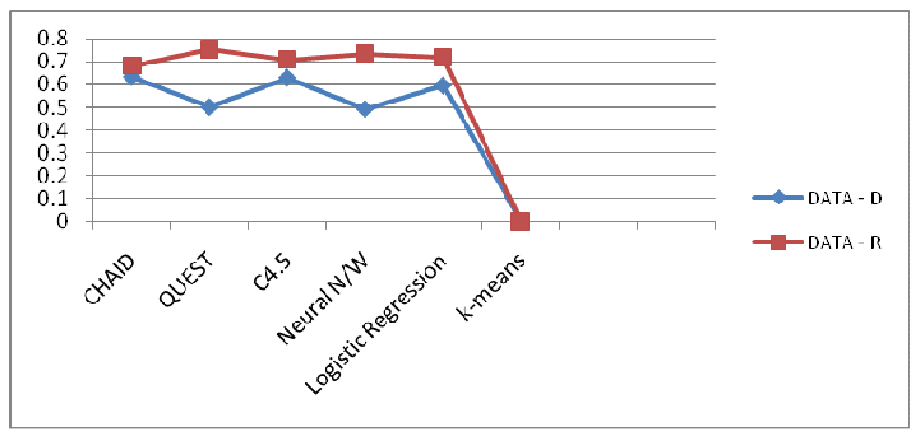

Figure5. Classification Index on Vote data set.

Figure (4-5) describes the classification index on Mushroom and Vote data sets for different classifiers. In these figures, neural network shows the highest classification index. Logistic regression is next to neural network. Decision tree based algorithms occupy the third position. SVM and SVM-ABoostM1 classification index is of the order of decision trees.

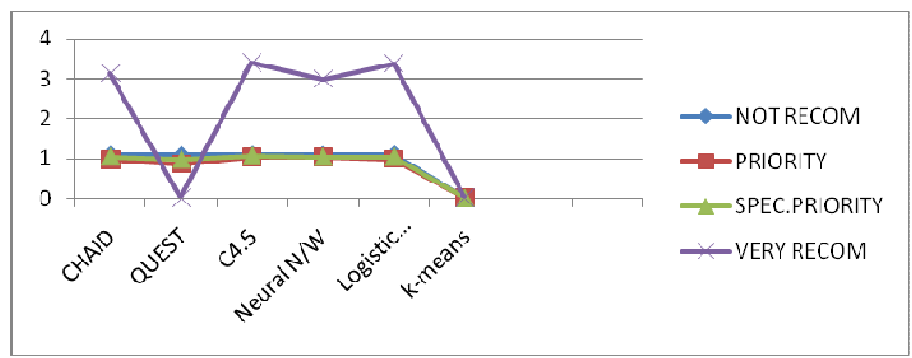

Figure6. Classification Index on Nursery data set.

In figure 6, algorithms show the variable classification index due to large distinct value of class attribute in nursery data set.

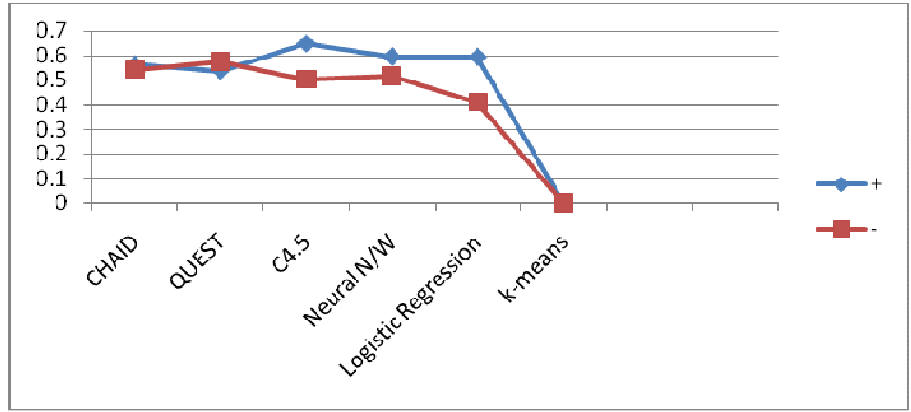

Figure7. Classification Index on Credit data set.

In figure 7, C4.5 shows the highest classification index. SVM and SVM-ABoostM1 classification index is of the order of decision trees. Other algorithms show their classification index with little differences.

## 4.5. Comprehensibility

Table 6 and figure (8-11) show the comprehensibility on Mushroom, Vote, Nursery and Credit data sets for all algorithms.

Table6. Comprehensibility on Mushroom, Vote, Nursery and Credit datasets.

| Data Set Classifier | Mushroom | | Vote | | Nursery | | Credit | |
|---|---|---|---|---|---|---|---|---|
| | Leaf Node | Depth | Leaf Node | Depth | Leaf Node | Depth | Leaf Node | Depth |
| CHAID | 5 | 2 | 5 | 3 | 21 | 5 | 9 | 5 |
| QUEST | 4 | 3 | 3 | 2 | 15 | 4 | 2 | 1 |
| C4.5 | 13 | 5 | 2 | 1 | 50 | 7 | 6 | 4 |
| Neural N/W | Nil | Nil | Nil | Nil | Nil | Nil | Nil | Nil |
| Logistic Regression | Nil | Nil | Nil | Nil | Nil | Nil | Nil | Nil |
| k-means | Nil | Nil | Nil | Nil | Nil | Nil | Nil | Nil |
| Genetic Algorithm | 6 | 4 | 8 | 6 | 4 | 2 | 5 | 3 |
| SVM | Nil | Nil | Nil | Nil | Nil | Nil | Nil | Nil |
| SVM-ABoostM1 | Nil | Nil | Nil | Nil | Nil | Nil | Nil | Nil |

Figure 8 given below describes the comprehensibility on Mushroom data set. Genetic algorithm and QUEST show the highest comprehensibility due to lesser number of leaf nodes and depth. k-means, logistic regression, neural network SVM and SVM-ABoostM1 show nil comprehensibility.

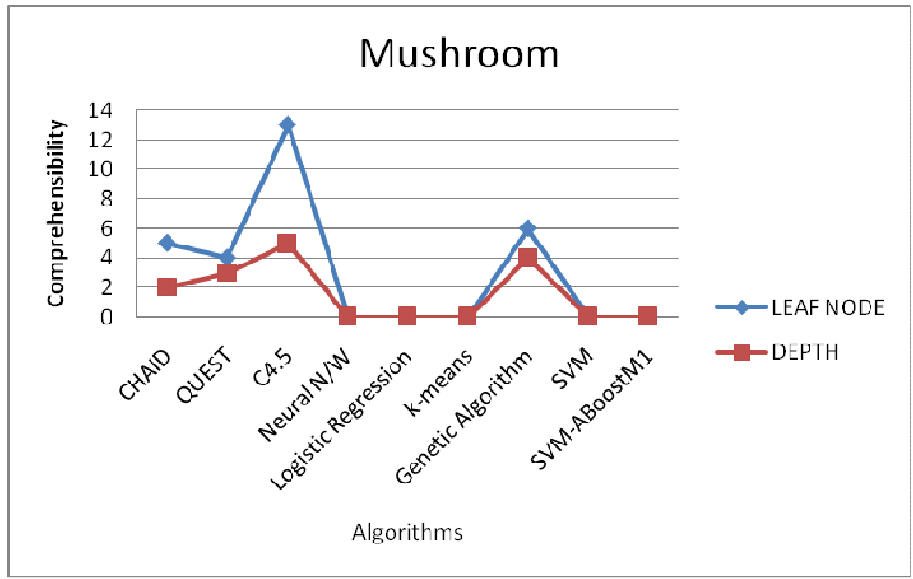

Figure8. Comprehensibility on Mushroom data set.

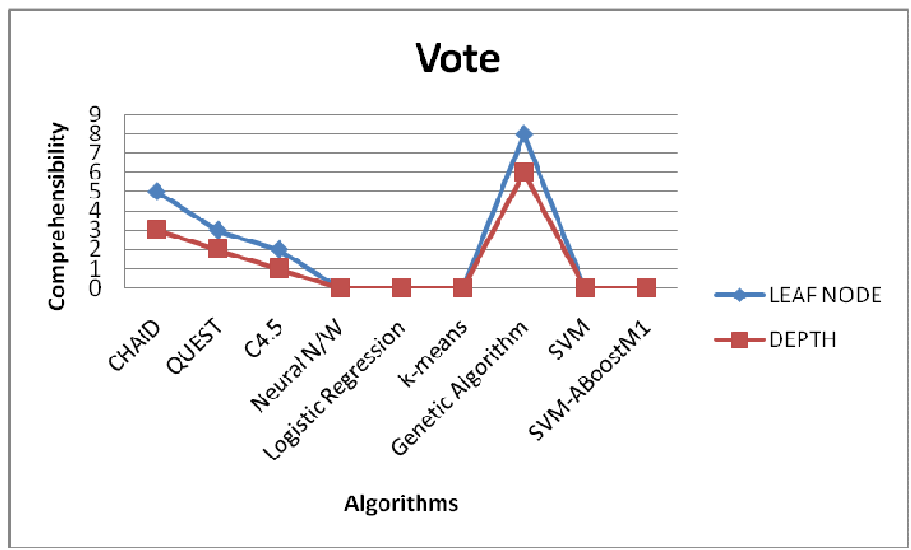

Figure9. Comprehensibility on Vote data set.

In figure 9, QUEST shows highest comprehensibility. k-means, logistic regression, SVM, SVM-ABoostM1 and neural network show nil comprehensibility. C4.5 and genetic algorithm show almost same comprehensibility.

In figure 10 given below, QUEST and Genetic algorithm show the highest predictive accuracy. k-means, logistic regression, neural network, SVM and SVM-ABoostM1 show nil comprehensibility. C4.5 is on the second last position and CHAID is on the last one.

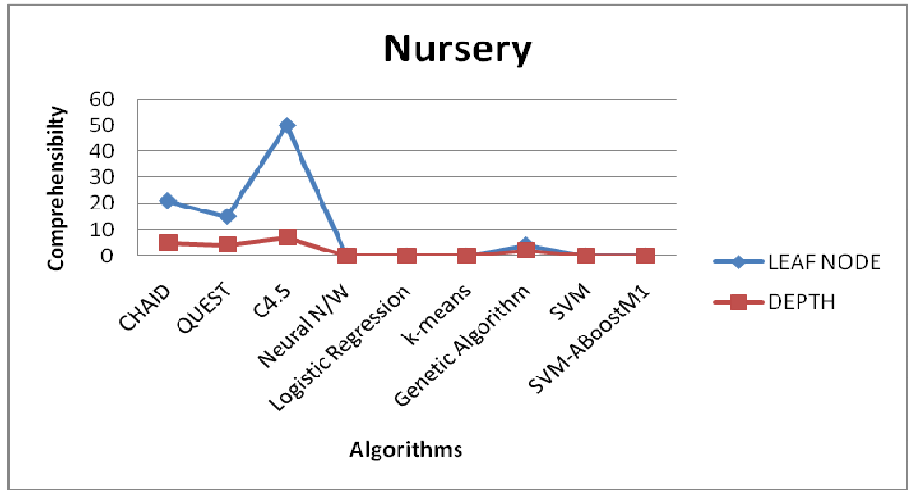

Figure10. Comprehensibility on Nursery data set.

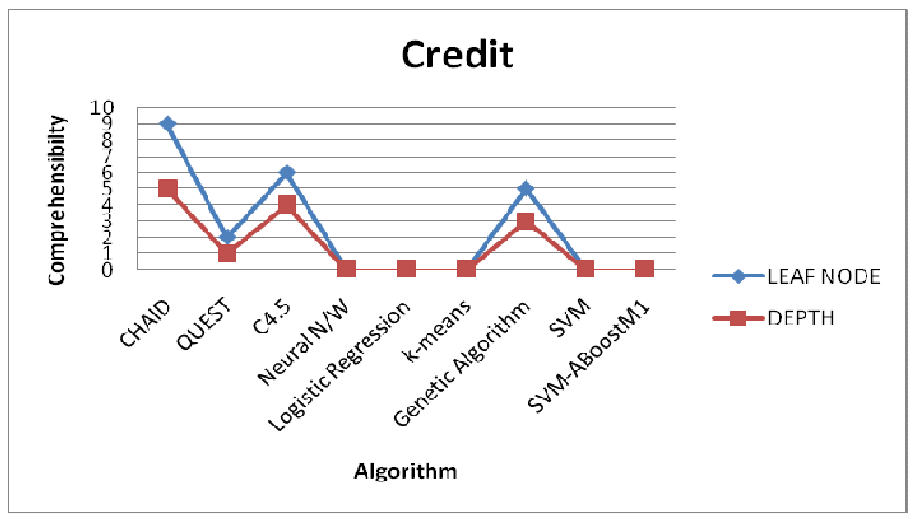

Figure11. Comprehensibility on Credit data set.

In figure 11, Genetic algorithm and QUEST show the highest predictive accuracy. C4.5 and CHAID show the worse comprehensibility comparatively. k-means, logistic regression, neural network, SVM and SVM-ABoostM1 show nil comprehensibility.

## 5. CONCLUSION

SVM, SVM-ABoostM1, Genetic algorithm and C 4.5 show the constant and maximum predictive accuracy independent of data set size and domain. Selector of classification algorithms should keep in mind that SVM doesn't perform well when the class attribute in a data set consists of large number of distinct values. Hence we can say that these algorithms are also better in terms of error rate with the condition described earlier. Training time is significant in case of large datasets. SVM, C 4.5, CHAID, QUEST and k-means are fastest algorithms. Logistic regression placed on second rank. Neural network and genetic algorithm are on the second last position. The former is slower due to back propagation execution process and later is the slower due to its chromosomal processing nature.SVM-ABoostM1 is slowest one due to boosting process execution nature.

Genetic algorithm is better in terms of comprehensibility as it is independent of data set size. K-Mean, neural network learning and logistic regression have no comprehensibility. Decision Tree (QUEST) is better in term of comprehensibility as like genetic algorithm, it is also independent of dataset size. QUEST is better in terms of classification index. k-means, neural network, logistic regression, SVM and SVMABoostM1 show nil comprehensibility.

Genetic algorithm is the first choice when predictive accuracy and comprehensibility are the selection criterion and decision tree(C 5.0) is the first choice when training time is a selection criterion. SVM is the first choice in terms of predictive accuracy and training time. Boosting for all algorithms is suggested but time required is too much large.

Decision makers should make tradeoff between various parameters and conditions described in this work to purchase data mining product for their decision support systems.

## AUTHORS

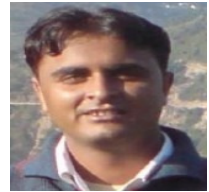

***Pardeep Kumar*** received his M.Tech (CSE) degree from Guru Jambheshwar University of Science & Technology, Hisar, Haryana, India. He received his B.Tech (IT) degree from Kurukshetra University, Haryana, India. Presently he is working as Senior Lecturer in department of CSE & IT at Jaypee University of Information Technology, at, Solan (H.P), India. His areas of interest include Machine Learning, Data Mining, Artificial Intelligence, Decision Support Systems and fusion of Data Mining with Image Processing

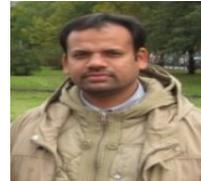

***Dr. Nitin*** received his Ph.D. (Computer Science & Engineering) degree from Jaypee University of Information Technology, Waknaghat, Solan (H.P), India Currently he is working as Associate Professor in the department of CSE & IT, Jaypee University of Information Technology, Waknaghat, Solan (H.P), India. He was Visiting Professor, University of Nebraska at Omaha, Omaha, USA (August 2010 to November 2010). His areas of interest include Parallel, Grid and Distributed Computing, Parallel Computing Algorithms, Super Computing etc. Currently he is the active reviewer of various international journals of repute published by IEEE, Springer, Elsevier and many more.

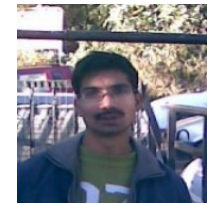

***Dr Vivek Kumar Sehgal*** received his Ph.D. (Computer Science & Engineering) degree from Uttarakhand Technical University, Dehradun, India. Currently he is working as Assistant Professor in CSE & IT department at Jaypee University of Information Technology, Waknaghat Solan (H.P), India. His areas of interest are Embedded Processor Architecture, Hardware Software Co-design, Smart Sensors and Systems-on-Chip and Networks-on-Chip, Machine learning etc.

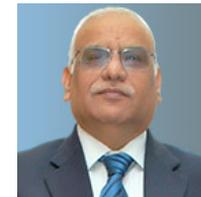

***Prof. Durg Singh Chauhan*** received his B.Sc Engg.(1972) in electrical engineering at I.T. B.H.U., M.E. (1978) at R.E.C. Tiruchirapalli (Madras University ) and Ph.D. (1986) at IIT/Delhi. He did his post doctoral work at Goddard space Flight Centre, Greenbelt Maryland. USA (1988- 91).His brilliant career brought him to teaching profession at Banaras Hindu University where he was Lecturer, Reader and then has been Professor till today. He has been director KNIT Sultanpur in 1999-2000 and founder vice Chancellor of U.P.Tech. University (2000- 2003-2006). Later on, he has served as Vice-Chancellor of Lovely Profession University (2006-07) and Jaypee University of Information Technology (2007-2009). Currently he has been serving as Vice- Chancellor of Uttarakhand Technical University for (2009-12) Tenure.